# Disclosure of Investment Advisor and Broker-Dealer Relationships: Impact on Comprehension and Decision Making

Xiaoqing Wan, Nichole R. Lighthall
Department of Psychology, University of Central Florida

Abstract

Recently enacted regulations aimed to enhance retail investors' understanding about different types of investment accounts. Toward this goal, the Securities and Exchange Commission (SEC) mandated that SEC-registered investment advisors and broker-dealers provide a brief relationship summary (Form CRS) to retail investors. The present study examines the impact of this regulation on investors and considers its market implications. The effects of Form CRS were evaluated based on three outcome variables: *perceived helpfulness*, *comprehension*, and *decision making*. The study also examined whether personal characteristics, such as investment experience, influenced the disclosure's impact on decision making. Results indicated that participants perceived the disclosure as helpful and it significantly enhanced comprehension about the two types of investment accounts. Critically, participants also showed increased preference and choice for broker-dealers after the disclosure. Increased preference for broker-dealers was associated with greater investment experience, greater comprehension gains, and access to more information from a longer disclosure. These findings suggest that Form CRS may promote informed decision making among retail investors while simultaneously increasing the selection of broker-dealer accounts.

## Introduction

The current study assessed the efficacy of Form Client Relationship Summary (Form CRS), a relatively new financial disclosure mandated by the U.S. Securities and Exchange Commission (SEC). Form CRS was created in 2018 to address the problem that many American investors do not understand the differences between broker-dealers and investment advisors (Hung et al. 2008; Securities and Exchange Commission 2018a), which was a problem identified following the 2007 – 2008 financial crisis. When investors open a brokerage account, they generally make their own investment decisions, monitor their own accounts, and pay a commission per trade. In contrast, an investment advisor is a fiduciary who typically charges a fee based on a percentage of the asset. Failing to understand these distinctions may result in investors entrusting their life savings to a financial service provider who is not aligned with their needs and preferences. To mitigate this problem, Form CRS was mandated in 2019 and to be disseminated by mid-2020 to all lay investors (Securities and Exchange Commission 2019).

Given the previous finding that ~90% of investors hold a favorable impression of Form CRS (Hung et al. 2018), we hypothesized that participants would perceive Form CRS as helpful. However, from the limited evidence available on the efficacy of this disclosure, it is not clear whether investors will pay attention to and gain information from the disclosure. It is also unclear whether changes in comprehension will induce changes in decision making. Since previous research on disclosure efficacy showed highly mixed findings (Ho et al. 2019), it remains an empirical question whether disclosures like Form CRS can help investors gain relevant knowledge, and whether this knowledge can in turn influence decision making.

**Determining Disclosure Efficacy**

From creating an online account to getting medical treatment, the presence of disclosure forms is ubiquitous. Mandated disclosures reveal facts and caveats about products and services. They disclose consumers' rights as well as conflicts of interest. In individual states, there are several hundred mandated disclosures (Schneider and Ben-Shahar 2010), which vary in length from just a few words to dozens of pages. Given the wide implementations of disclosures but conflicting results surrounding their efficacy, researchers have called for studies to understand "when, why, and how" disclosures benefit consumers (Loewenstein et al. 2014).

Given decades of research on disclosure efficacy, can we predict if a given disclosure will yield the targeted benefits for consumers? Or if it will exert any influence at all? These questions cannot be answered by consulting past disclosure research alone, as the literature includes both mixed results and research approaches. Benefits and null effects are both heavily documented in research on disclosure effectiveness (Ho et al. 2019). Unintended, harmful consequences are more often observed in situations involving social interactions (Cain et al. 2010; LeBoeuf et al. 2016; Sah et al. 2013), during which the discloser can bias a consumer's interpretation.

In terms of product warning labels, an accumulation of studies over the years yielded numerous review papers (e.g., Stewart & Martin, 1994; McCarthy et al., 1984) and two meta-analyses (Argo and Main 2004; Cox III et al. 1997). Collectively, they examined several hundred studies, but could not reach a unifying conclusion about whether disclosures produced more optimal behavior. In the current study, we seek to understand whether and how a recently mandated disclosure – Form CRS – impacts consumer choice. We do so by integrating approaches from learning and memory research with those of judgment and decision-making research.

**Measures of Disclosure Efficacy**

There are different ways to measure the effectiveness of a disclosure (Stewart and Martin 1994). We broadly organize them into three categories: (1) subjective impressions of whether people feel a disclosure is helpful and useful (e.g. Ho & Wong, 2004; Hung, Carman, Cerully, Dominitz, & Edwards, 2018; Oh, Nguyen, & Patrick, 2016. (2) More objective measures of attention and comprehension gains (e.g. Obar & Oeldorf-Hirsch, 2018; Felt et al., 2012; Goldhaber & DeTurck, 1988). (3) Disclosure-targeted changes in behavior and decision making (e.g. Choi, Laibson, & Madrian, 2009; Bertrand & Morse, 2011; Beshears, Choi, Laibson, & Madrian, 2009). Critically, these measurements capture different aspects of disclosure efficacy and it is presently unclear whether Form CRS can produce comprehension gains that influence decision making.

      Within the first category, studies that assess self-reported feelings of efficacy typically find that people perceive disclosures to be helpful (e.g. da Silva Nogueira & Jorge, 2017; Tooley & Hooks, 2010). However, this may be because people believe that having relevant information *should be* helpful (i.e., placebo effect). In a review of medical disclosures, Lemaire (2006) found that patients were highly satisfied with the information provided, but they subsequently showed poor recall of the operations they received, the diagnoses they had, and the risks they were exposed to. Thus, people may have a positive perception of disclosure efficacy even if they do not successfully retrieve the disclosure information they received. Indeed, insights from psychological science indicate that subjective perceptions can be a poor predictor of actual performance (Dunning 2011). For this reason, researchers have opined that conclusions about disclosure efficacy should not be solely based on subjective evaluations (Cox III et al. 1997).

      In the second category, researchers often use reading time, recognition, and recall to index disclosure efficacy. Compared to subjective feelings, comprehension-based outcome measures are more objective. Comprehension questions also serve to verify if participants are paying attention. For example, in a study about privacy disclosures, Obar and Oeldorf-Hirsch (2018) embedded "gotcha clauses" in text and measured how much time participants spent on reading the document. It was found that participants spent about one minute reading a disclosure that should have taken ~30 minutes to complete. Moreover, 98% of their participants agreed to the outlandish gotcha clause of giving away their first-born child to the website developer. In the disclosure literature, comprehension is typically assessed using multiple choice or short answers that tap the content of the disclosure (e.g. Wang, 2012). Despite the benefit of being more objective, comprehension may or may not translate into actual decision-making benefits (Stewart and Martin 1994).

The third category is arguably the most relevant index of disclosure efficacy, as the basic goal of information campaigns is to alter decision behavior. Decision behavior measures have been implemented both in the lab and the field. For example, researchers have studied whether disclosures can persuade people to (1) avoid costly payday loans (Bertrand and Morse 2011) and index funds (Choi et al. 2009), (2) reduce amassing credit card fees (Agarwal et al. 2014), and (3) to avoid visiting restaurants with poor hygiene (Jin and Leslie 2003). In general, these studies tend to find null results or small benefits from disclosures. A notable exception is the restaurant hygiene disclosure, which was reported to yield large reductions in foodborne illness hospitalizations (Jin and Leslie 2003). However, this effect did not survive re-analysis which included surrounding counties as control groups (Ho et al. 2019).

Overall, the field has presented multiple ways to index disclosure efficacy. We leverage these insights from previous research in the current study. Specifically, we critically evaluated the effects of Form CRS by measuring its impact on (1) perceived helpfulness, (2) comprehension, and (3) decision making. A second aim of the study is to examine which content areas from Form CRS influenced decision making, and in which direction.

**Additional Factors Influencing Disclosure Efficacy**

In addition to evaluating the disclosure efficacy, we examined whether Form CRS would benefit certain individuals more than others. It has been argued that learning-by-reading is different from learning-by-doing (Pfeffer and Sutton 2001). The latter deals with practice, experience, and trial and error, whereas the former is a less sophisticated kind of learning (Schneider and Ben-Shahar 2010). It is possible that participants with more investment experience can stand to benefit more from a disclosure, because they know what to look for, and they can efficiently extract relevant information (Ettenson et al. 1987; Grether et al. 1985). To investigate the effect of experience on disclosure efficacy, we recruited participants across the lifespan with different levels of investment experience.

Finally, we experimentally manipulated disclosure length to examine the role of information reduction in disclosure efficacy. An assumption from the literature on information campaigns is that, since people do not and cannot process too much information at once, simplified information may be more effective. Following this assumption, a "plain English movement" was started in the 1970s to cut back legalese (Securities and Exchange Commission 1997). Similar efforts devoted to simplifying disclosures have been made since the Truth-in-Lending Simplification and Reform Act of 1980 (O'Connor 1981).

Despite these efforts, whether simplification makes for better disclosures remains a thorny question. Studies that did address this query returned conflicting results. For example, while a simplified disclosure nudged eligible participants to claim tax credit more often compared to a longer version, (Bhargava and Manoli 2012), another simplified disclosure could not improve participants' mutual fund choices (Beshears et al. 2009). Schneider and Ben-Shahar (2010) criticized disclosure simplification for several reasons: a complicated issue cannot be simplified without losing information, thus simplification goes against total transparency. Also, to shorten a disclosure, one does not always have good criteria for what information should be omitted. To empirically investigate the effect of information reduction on disclosure efficacy, the present study compared the differences in comprehension between a simplified, shorter Form CRS and the original, longer SEC version. Toward this goal, we leveraged a shorter version of Form CRS that was developed from a previous qualitative study with iterative testing via structured interviews (Kleimann Communication Group 2018).

        To address these aims, the present study empirically tested the effectiveness of Form CRS, using an adult lifespan sample. We examined the effects of Form CRS on three outcome measures: perceived helpfulness, comprehension, and decision making. To understand whether the disclosure influences individuals differently, we examined the effects of age, vocabulary, income, education, and investment experience on decision change. Finally, we examined what specific content from the disclosure influenced participants' decision making.

## Methods

### Participants

Using MTurk, 352 younger (age 19 - 35), 225 middle-aged (age 36 - 59), and 119 older (age 60 - 78) adult participants were recruited. The MTurk premium qualification feature was used in recruiting older adults to ensure a comparable sample size for this group. Eligibility requirements included that participants primarily resided in the U.S., were fluent in English, and did not report any neurological conditions that would affect their cognitive function.

        For data quality control, participants were excluded from analysis if they met one of the following criteria: (1) failing catch trials (e.g., "please choose the 4th response option to demonstrate that you are paying attention"); (2) providing incomprehensible answers (e.g., typing non-words); (3) failing to follow instructions (e.g., typing random words); (4) the duration of their overall participation was three standard deviations away from the mean. The final sample consisted of 618 participants (Table 1). There were no differences in the outcome variables by gender at baseline ($ps > .62$). Upon completion of the study, participants received a participation stipend of $5 through the MTurk compensation system. This study was approved by the university's Institutional Review Board, and all participants gave informed consent.

Table 1. Sample demographics

|  | **Younger Adults** | **Middle-Aged Adults** | **Older Adults** |
|---|---|---|---|
| *n* | 282 | 218 | 114 |
| Female | 32.9% | 52.6% | 61.4% |
| Age | 29.3 (3.7) Range 19 - 35 | 46.3 (7.9) Range 36 - 59 | 64.9 (3.6) Range 60 – 78 |
| Years of Education | 14.2 (1.3) | 14.2 (1.2) | 14.2 (1.4) |
| Mean Household Income | $50,000 to $74,999 | $50,000 to $74,999 | $50,000 to $74,999 |
| High Investment Experience | 48.6% | 59.2% | 60.5% |
| Vocabulary Score | 32.1/40 (4.7) | 34.4/40 (3.7) | 35.5/40 (3.7) |

Notes: Mean (SD) are shown for continuous variables. Four participants did not report age and were not counted in the row "*n*".

**Stimuli**

The disclosure form used in the study was Form CRS, which was mandated in 2019 to be provided to all retail investors who open an account with a broker-dealer or an investment advisor. Participants were randomly assigned one of two versions of Form CRS. The longer version (Appendix A) was developed by the Securities and Exchange Commission (Securities and Exchange Commission 2018b). The shorter version (Appendix B) was developed by Kleimann Communication Group with support from AARP (Kleimann Communication Group 2018). For simplicity, we refer to the two disclosure conditions as "longer" and "shorter" versions. This design allowed us to investigate the effect of length on disclosure efficacy, with the shorter version having a word count reduction of 44.8% compared to the longer version. Both versions were designed to explain important differences between broker-dealers and investment advisors, and they aimed to help investors make decisions that are consistent with their values and needs. On average, participants spent significantly more time reviewing the longer version ($M = 6.1$ min; $SD = 4.6$ min) compared with the shorter version ($M = 4.2$ min; $SD = 3.1$ min; $F(1, 264) = 15.39$, $p < .001$). Additionally, older adults spent more time browsing the disclosure ($M = 7.6$ min; $SD = 3.7$ min) compared with younger adults ($M = 2.6$ min; $SD = 3.4$ min) across disclosure versions ($F(1, 264) = 138.16$, $p < .001$), with no interaction between age group and disclosure version ($p = .11$). While reading the disclosure, participants could freely advance to the next section or return to previous sections until satisfied.

**Procedure**

The study was completed online via QualtricsXM experimental software (https://www.qualtrics.com/). Participants first provided demographic and socio-economic information. They then reported their investment experience (e.g. In the past, have you had any investments in stocks or mutual funds?) via a questionnaire adapted from Hung and colleagues (2018). The investment experience questionnaire consisted of six multiple-choice questions and one short answer text box for reporting on types of personal investment holdings. Participants answering yes to three or more multiple-choice questions were coded as having "more experience." Those with less than three affirmative responses were coded as having "less experience."

Two outcome variables – comprehension and decision making – were assessed twice: once before and once after participants received the disclosure. Ratings of perceived helpfulness were then reported on a scale of 0 (*extremely unhelpful*) to 100 (*extremely helpful*). Participants also reported their preferences for three investment features (Appendix C; results not included in this paper). Finally, participants completed the Shipley institute of living scale: vocabulary (Shipley and Zachary 1986) and additional cognitive tasks (results not included in this paper).

**Comprehension**

We assessed participants' comprehension of financial service providers before and after they received the disclosure to determine whether they gained new information. The change in comprehension from pre- to post- intervention reflects learned information from the disclosure. The comprehension questionnaire[1] consisted of eight fact-based questions (Table 2). Answers were derived based on direct quotes from Form CRS (see Appendix D). The wording of the comprehension questions was carefully chosen to avoid bias towards either version. In addition, the answers to the questions could be found in both versions of the disclosure.

---

[1] The comprehension questionnaire also included a question about information provided by the investor.gov website. This question was excluded from our analysis because participants were not given the opportunity to navigate the website to learn more about financial service providers during the experiments.

Table 2. Comprehension questions and answers derived from Form Client Relationship Summary

| | Questions | Answers |
|---|---|---|
| 1 | Which financial professionals have a transaction-based relationship with you? | Broker-dealer |
| 2 | Which financial professionals offer trading recommendations, but ask you to make the final decision? | Both |
| 3 | Which financial professionals are held to a fiduciary standard? | Investment Advisor |
| 4 | Which financial professionals charge fees primarily based on the amount of your assets? | Investment Advisor |
| 5 | Which financial professionals are paid primarily from commissions? | Broker-dealers |
| 6 | Which financial professionals have an incentive to sell investment products offered by companies with whom they have a relationship? | Both |
| 7 | Which financial professionals have an incentive to sell investment products that will result in higher revenue or extra income for them? | Both |
| 8 | Which financial professionals monitor your accounts on an ongoing basis? | Both, Investment advisor* |

*For question 8, the longer version and the shorter version disagree regarding whether broker-dealers can monitor a client's account. "Both" was coded as the answer for the longer version, and "Investment advisor" the answer for the shorter version. Question 8 was analyzed separately in Appendix E.

Regarding which financial professional monitors their client's account on an ongoing basis (question 8), the two disclosure versions disagreed on the correct answer (see Appendix D). Subsequently, this question was analyzed separately. Regardless of version, participants generally believed that investment advisors, as opposed to broker-dealers, assume the responsibility of monitoring accounts. This belief was further strengthened post disclosure across both versions (see Appendix E).

We provided a slide bar for participants to indicate both the answer and the level of self-reported confidence (Appendix F). The correct answer to the comprehension questions was either "broker-dealer," "investment advisor," or "both." Confidence was assessed to reduce the effect of guessing correctly by chance and to reduce ceiling effects that may occur if participants came in with high baseline comprehension (Stewart and Martin 1994). A score of 100 reflected correct responses at the highest confidence level. A score of 0 reflected incorrect responses at the highest level of confidence.

The slide bar allowed for a range of responses from the highest confidence for "broker-dealers" to the highest confidence for "investment advisors." Responses in the center of the slide bar indicated an answer of "both." For example, if a participant moved the slide bar all the way towards "broker-dealer" for question 1, they would get 100 for this question; if moved all the way towards "investment advisor," then a score of 0; if moved exactly to the center, then a score of 50, since "both" contained "broker-dealer" and was half correct.

Here is another example: for question 2, if participants moved the slide bar to "both," the correct answer, they would get 100. The lowest score one can get when the correct answer is "both" is 50, since answering either "broker-dealer" or "investment advisor" is half correct.

Finally, comprehension scores were averaged based on questions 1-7, calculated separately for baseline comprehension and follow-up comprehension. Comprehension change was defined as follow-up comprehension minus baseline comprehension.

**Decision Making**

Decision making was assessed by two experimental phases (Appendix G). First, during the decision phase, participants were given a hypothetical scenario followed by a binary choice:

> "You have money that you want to invest and you do not currently have any investment accounts. You have contacted an investment firm that offers two types of services: broker-dealer services and investment advisor services. Which services do you choose?"

Participants then made a forced-choice, binary decision between broker-dealers and investment advisors. After that, participants reported the confidence of their decision on a 0 to 100 slide bar, with higher values indicating a more confident decision.

Repeating these phases both before and after participants received the disclosure yielded two continuous measures: baseline decision and follow-up decision. They were calculated as follows: participants who chose broker-dealers (or investment advisors) were assigned a value of -1 (or 1); this was then multiplied by the confidence, yielding a range of -100 (the participant chose broker-dealers with the highest confidence) to 100 (the participant chose investment advisors with the highest confidence).

Finally, the continuous decision change was calculated as follow-up decision minus baseline decision. For example, we have a participant who initially chose broker-dealers with the highest confidence. After reading the disclosure, they were convinced to switch to investment advisors with the highest confidence. In this case, the continuous decision change would be 100 – (-100) = 200.

## Statistical Analysis

All statistical analyses were completed in R version 4.0.2 (R Core Team 2020). To test the efficacy of the disclosure, separate models were performed for the three outcome measures (i.e. perceived disclosure helpfulness, comprehension, and decision making). One sample t-test was used to examine whether participants perceived the disclosure as helpful. Linear regression was used to determine whether participants experienced comprehension change; the outcome variable was decision change and age, version, and their interactions were independent variables. Mean, standard deviation, and Cohen's d were calculated to determine whether each comprehension question elicited (1) better-than-chance baseline performance, and (2) significant improvements or decrements post disclosure.

Decision making was indexed in two ways: (1) a forced-choice, binary decision between broker-dealers and investment advisors. (2) Continuous decision change reflected how participants changed their decision post disclosure and their confidence. To assess how participants' binary decision between investment advisors and broker-dealers changed post disclosure, one proportion z test was performed. Next, a linear regression was performed with continuous decision change as the outcome variable. The independent variables were age, version, age * version, vocabulary, education, income, experience, baseline decision, baseline comprehension, and comprehension change.

Finally, we assessed how comprehension change from individual questions influenced the continuous measure of decision change. For example, the disclosure stated that an investment advisor is a fiduciary while broker-dealers is not; how does this information impact participants' decision regarding choosing a financial service provider? To address this, a linear regression was performed with continuous decision change as the outcome variable, and with comprehension change separately from individual questions entered as independent variables.

## Results

**Participants perceived the disclosure as helpful**

Consistent with prior findings ((Kleimann Communication Group 2018), participants overwhelmingly perceived the disclosure as helpful ($t(606) = 49.37$, $p < .001$), with an average rating of ~81 out of 100 (95% *CI*: 79.3, 81.7) compared to the neutral rating of 50. This effect did not differ by age group ($p = .97$) or disclosure version ($p = .82$).

**Comprehension improved post disclosure**

After reviewing Form CRS, participants' comprehension significantly improved from baseline (intercept estimate = 5.16, SE = 1.96, t = 2.64, p = .009). Age (p = .12), version (p = .26), and their interaction (p = .57) had no effect on comprehension change.

When each comprehension question was examined separately (Table 3), results showed that participants entered the study with a level of baseline comprehension that was significantly better than chance. Post disclosure, comprehension performance was significantly greater for all questions except one (Question 2), which showed a small but significant decrement. Question 2 asked, "Which financial professionals offer trading recommendations, but ask you to make the final decision?" The correct answer ("both") may have been ambiguous to participants; although investment advisors can defer to the clients to make the final decisions, they can also take over this responsibility if the client chooses.

Table 3. Baseline comprehension and comprehension change by individual comprehension questions

| Question | Baseline Comprehension | | Comprehension Change | |
| --- | --- | --- | --- | --- |
| | M(SD) | Cohen's d | M(SD) | Cohen's d |
| 1 | 62.97(28.23)*** | 0.46 | 13.42(29.86)*** | 0.45 |
| 2 | 73.16(17.26)*** | 1.34 | -2.09(22.95)* | 0.09 |
| 3 | 55.62(24.18)*** | 0.23 | 17.33(30.44))*** | 0.57 |
| 4 | 62.21(27.47)*** | 0.44 | 16.95(33.82)*** | 0.50 |
| 5 | 63.08(29.22)*** | 0.45 | 10.11(33.54)*** | 0.30 |
| 6 | 76.79(17.97)*** | 1.49 | 8.63(22.64)*** | 0.38 |
| 7 | 76.20(17.73)*** | 1.48 | 6.21(23.26)*** | 0.27 |

Note: mu = 50 for baseline comprehension; mu = 0 for comprehension change. $p < .05$*; $p < .001$***.

**Decisions changed towards broker-dealers post disclosure**

Results for the forced-choice, binary decision[2] between broker-dealers and investment advisors are presented in Table 4. Most participants initially chose investment advisors as their service provider, but there was a systematic shift towards broker-dealers after viewing Form CRS. The overall pattern of data showed that 21.5% of participants categorically switched from investment advisors to broker-dealers after reviewing the disclosure.

---

[2] Not all participants were confident about their binary decision. At baseline, 15 participants (2.5% of total sample) indicated zero strength for their decisions. Post disclosure, only two participants reported zero strength and the rest were evenly split between investment advisors (7) and broker-dealers (6). Exclusion of these participants did not alter study findings.

Table 4. The number (percentages) of participants who chose investment advisors or broker-dealers

|  | Pre disclosure | Post disclosure | |
|---|---|---|---|
| Chose investment advisors | 480 (77.7%) | 380 (61.5%) | |
| Chose broker-dealers | 138 (22.3%) | 238 (38.5%) | |
| Investment advisors to broker-dealers ratio | 3.48 | 1.60 | χ2 = 92.37*** |

Note: Participants were given a two-alternative, forced choice between investment advisors and broker-dealers. One proportion z test was used to determine whether the ratio between investment advisors and broker-dealers changed post disclosure. p < .001 ***.

Next, the continuous decision change was examined as the outcome variable. Results from linear regression (Table 5) revealed potential effects driving the shift towards broker-dealers. Four types of participants were more likely to shift towards broker-dealers: (1) participants who had more investment experience; (2) participants who had more comprehension gains; (3) participants who received the longer version of disclosure; (4) participants who felt strongly about choosing investment advisors at baseline.

Table 5. Linear regression effects on continuous decision change

|  | β | SE | t value | p value |
|---|---|---|---|---|
| (intercept) | -6.92 | 37.88 | -0.18 | .86 |
| Age | -3.47 | 3.81 | -0.91 | .36 |
| Version: longer | -37.84 | 5.12 | -7.40 | < .001*** |
| Age * Version: longer | -7.80 | 5.10 | -1.53 | .13 |
| Vocabulary | 0.64 | 0.55 | 1.16 | .25 |
| Education | 2.50 | 2.18 | 1.15 | .25 |
| Income | -0.10 | 0.07 | -1.49 | .14 |
| Experience: more | -13.74 | 5.57 | -2.47 | .01* |
| Baseline Decision | -0.39 | 0.05 | -8.63 | <.001*** |
| Baseline Comprehension | -0.17 | 0.37 | -0.47 | .64 |
| Comprehension Change | -0.91 | 0.28 | -3.29 | .001** |
| Adjusted R squared |  |  | 0.22 |  |

Note: negative coefficients indicate a change towards broker-dealers. p < .05*. p < .01**. p < .001***.

To further examine decision making, we analyzed whether participants' switching behavior was influenced by version, investment experience, or comprehension change.

As participants updated their decisions regarding whether to choose investment advisors or broker-dealers after reviewing the disclosure, there were three possible switching behaviors: (1) participants could switch from investment advisors to broker-dealers, (2) from broker-dealers to investment advisors, or (3) no switch. Most participants (73.1%) made no switch and stayed committed to their original choice. Critically, those who changed their decision mostly switched from investment advisors to broker-dealers (21.5%). Only a small portion of participants (5.3%) switched from broker-dealers to investment advisors. Furthermore, the chi square test of independence revealed that participants who switched their decision from investment advisors to broker-dealers were more likely to have received the longer version of the disclosure and/or have improved their comprehension.

Table 6. Decision switching by disclosure version, investor experience, and comprehension change

| Decision Switching | Version | | Experience | | Comprehension change | | Overall |
|---|---|---|---|---|---|---|---|
| | Shorter | Longer | Less | More | Declined/ no change | Improved | |
| Investment advisors to broker-dealers | 33 (24.8%) | 100 (75.2%) | 58 (43.6%) | 75 (56.4%) | 21 (15.8%) | 112 (84.2%) | 133 (100%) |
| Broker-dealers to investment advisors | 22 (66.7%) | 11 (33.3%) | 17 (51.5%) | 16 (48.5%) | 5 (15.2%) | 28 (84.8%) | 33 (100%) |
| No switch | 255 (56.4%) | 197 (43.6%) | 204 (45.1%) | 248 (54.9%) | 127 (28.1%) | 325 (71.9%) | 452 (100%) |
| | $\chi^2 = 44.86$*** | | $\chi^2 = .67$ | | $\chi^2 = 10.08$** | | |

Note: The number of participants (percentages) who switched their decisions or made no switch are shown in each cell. $p < .05$*. $p < .01$**. $p < .001$***.

Since comprehension change was an important predictor of decisions, we examined how individual comprehension questions influenced continuous decision change. Results revealed that four content areas were significantly associated with decision change (Table 7). First, results indicated that as participants learned that investment advisors have fiduciary duties, decisions shifted towards investment advisors. However, this effect was weaker compared to effects associated with the other content areas. Decisions shifted towards broker-dealers after participants learned about three facts: (1) investment advisors charges asset-based fees, (2) both investment advisors and broker-dealers are incentivized to sell products offered by companies with whom they have a relationship, and (3) both investment advisors and broker-dealers are incentivized to sell investment products that will result in higher revenue or extra income for them. Thus, consistent with the overall findings, post-disclosure gains in comprehension were associated with increased choices for broker-dealers.

Table 7. The effects of change from individual comprehension questions on continuous decision change.

| Questions | β | SE | t value | p value |
|---|---|---|---|---|
| (intercept) | -4.20 | 3.69 | -1.14 | 0.26 |
| 1 | 0.00 | 0.09 | 0.00 | 1.00 |
| 2 | 0.15 | 0.12 | 1.25 | 0.21 |
| 3 | 0.19 | 0.09 | 2.06 | 0.04* |
| 4 | -0.26 | 0.09 | -3.06 | .002** |
| 5 | -0.14 | 0.08 | -1.66 | 0.10 |
| 6 | -0.53 | 0.13 | -3.97 | <.001*** |
| 7 | -0.37 | 0.13 | -2.85 | .005** |
| Adjusted R squared | | | 0.07 | |

**Discussion**

Guided by prior research on disclosure efficacy, the current study was the first to systematically assess the impact of Form CRS, a relatively new financial disclosure mandated by the SEC. This well-powered, randomized, controlled experiment evaluated the effects of Form CRS on three outcome variables: perceived helpfulness, comprehension, and decision making about different types of investment accounts. The study also assessed how personal characteristics influenced disclosure efficacy, and whether information reduction enhanced disclosure efficacy. The latter goal was accomplished by randomly assigning participants to either the longer, SEC version or an alternative, shorter version.

      Participants ranged from 18 to 90 years old and had varying levels of investment experience. For all participants, comprehension and decision making were assessed pre- and post-disclosure. Consistent with earlier findings, results showed that participants perceived the disclosure as helpful. Additionally, comprehension about the services offered by broker-dealers and investment advisors significantly improved after reviewing the disclosure, regardless of which disclosure version was received. For decision making, participants significantly shifted their preferences and choice from investment advisors to broker-dealers after receiving the disclosure. The increase in preferences for broker-dealers was more pronounced in four types of participants: those who had more investment experience; those who had more comprehension gains; those who received the longer version of disclosure; those who felt strongly about choosing investment advisors before receiving disclosure.

**Gains in comprehension**

Our first central finding indicated robust gains in comprehension regardless of disclosure versions. This suggests that Form CRS was able to effectively inform investors about the differences between broker-dealers and investment advisors. Participants showed large comprehension gains for the fact that investment advisors are held to a fiduciary duty and that they charge asset-based fees. Participants also showed moderate comprehension improvements for the following knowledge: (1) broker-dealers have a transaction-based relationship with clients; (2) broker-dealers are paid primarily from commissions; (3) financial professionals are incentivized to sell investment products offered by companies with whom they have a relationship. Small improvements were observed for the knowledge that both investment advisors and broker-dealers are incentivized to sell investment products that will result in higher revenue or extra income for them.

  In contrast to these comprehension improvements, there was a small decline in comprehension about which financial professional asks the client to make the final decisions. The selective decline in this question may have occurred because the answer (i.e., both broker-dealers and investment advisors) is somewhat ambiguous. That is, while investment advisors allow clients to make buy-and-sell decisions, they can also take over this responsibility if the client chooses.

  Additionally, we examined the possibility that the shorter version may induce greater benefits due to information reduction. The shorter disclosure was created based on findings from Kleimann Communication Group's research report (Kleimann Communication Group 2018), as well as consultation with SEC Chairman Jay Clayton, which was commissioned by AARP, Consumer Federation of America, and the Financial Planning Coalition. Since this shorter form was developed from research partially sponsored by the AARP, we investigated whether the shorter form may yield greater benefits for older adults in terms of comprehension. However, our results indicated that, regardless of age, the longer and shorter versions induced equivalent gains in perceived helpfulness and comprehension.

**Increased choice for broker-dealer**

Our second important finding indicated that reviewing the disclosure significantly increased participants' preference and choice for broker-dealers. Indeed, after receiving the disclosure, ~21% of all participants categorically changed their choice from investment advisors to broker-dealers. After receiving the disclosure, participants who initially had a stronger preference for investment advisors were more likely to shift towards broker-dealers. Thus, while the SEC did not explicitly design Form CRS to promote either investment advisors or broker-dealers (Securities and Exchange Commission 2019), our findings suggest that this mandated disclosure may increase the proportion of American investors choosing broker-dealers in the future.

  Correspondingly, post-disclosure gains in comprehension were almost exclusively associated with an increased preference for broker-dealers. As participants learned that investment advisors charge fees primarily based on the amount of their assets, decisions shifted towards broker-dealers. This shift may reflect the motivation to reduce cost, since asset-based fees are generally more expensive than commissions, provided that the investor does not trade often. In contrast, commissions charged by broker-dealers have been racing towards zero in recent years (Chang 2019).

Despite the overall shift towards broker-dealers, participants' decisions shifted towards investment advisors when they learned that investment advisors are held to a fiduciary standard – a duty of care and loyalty. However, this effect appeared to be eclipsed by the other information gained from the disclosure. For example, when participants learned that both financial professionals are incentivized to sell products that result in higher revenue for them, decisions shifted towards broker-dealers. Thus, Form CRS may help participants form the perception that although investment advisors are held to the highest standard of conduct, they are not free from conflicts of interest.

**Investor characteristics associated with disclosure effects**

The present study also highlighted investors' characteristics that can mediate the impact of Form CRS on decision making. After receiving the disclosure, those who had greater investment experience and more comprehension gains were likely to develop a stronger preference for broker-dealers. Shifts in preferences towards broker-dealers were also greater for those who received the longer, SEC version. Collectively, these findings indicate that a detailed disclosure is likely to strengthen preferences for broker-dealers among experienced investors, who are better equipped to extract meaningful information from a long disclosure. Such interpretations are consistent with prior studies that show experience can enable more efficient filtering of dense information (Gegenfurtner et al. 2011).

**Conclusions and implications**

The current experiment found that Form CRS significantly increased preferences and choices for broker-dealers over investment advisors. As Form CRS is mandated and widely implemented in the U.S., the financial industry may observe declines in investors choosing investment advisors and increases in broker-dealers. Participants' increased preference for broker-dealers (or decreased preferences for investment advisors) was likely influenced by learning the fact that investment advisors charge asset-based fees, which could be costlier for the average investor. Finally, this experiment suggested that the shifts toward broker-dealers were likely informed, rather than misguided decisions, since this change in preference was found among investors who had greater investment experience, greater comprehension gains, and access to more information from a longer, more detailed disclosure.

Appendix A. The longer, SEC version of Form CRS (participants read each section on a separate page)

# Relationship Summary

**ABC Financial and Wealth Management Services. Inc.
A Registered Investment Adviser and Broker-Dealer**

**Which Type of Account is Right for You – Brokerage, Investment Advisory or Both?**

There are different ways you can get help with your investments. You should carefully consider which types of accounts and services are right for you.

**Depending on your needs and investment objectives, we can provide you with services in a brokerage account, investment advisory account, or both at the same time.** This document gives you a summary of the types of services we provide and how you pay. Please ask us for more information. There are some suggested questions on the last page.

| *Broker-Dealer Services* **Brokerage Accounts** | *Investment Adviser Services* **Advisory Accounts** |
|---|---|
| **Types of Relationships and Services.** *Our accounts and services fall into two categories.* ||
| • If you open a brokerage account, you will pay us a ***transaction-based fee***, generally referred to as a commission, every time you buy or sell an investment.<br><br>• You may select investments or we may recommend investments for your account, but the ultimate investment decision for your investment strategy and the purchase or sale of investments will be yours.<br><br> • We can offer you additional services to assist you in developing and executing your investment strategy and monitoring the performance of your account but you might pay more. We will deliver account statements to you each quarter in paper or electronically.<br><br>• We offer a limited selection of investments. Other firms could offer a wider range of choices, some of which might have lower costs. | • If you open an advisory account, you will pay an on-going ***asset-based fee*** for our services.<br><br>• We will offer you advice on a regular basis. We will discuss your investment goals design with you a strategy to achieve your investment goals, and regularly monitor your account. We will contact you (by phone or e-mail) at least quarterly to discuss your portfolio.<br><br>• You can choose an account that allows us to buy and sell investments in your account without asking you in advance (a "***discretionary account***") or we may give you advice and you decide what investments to buy and sell (a "***non-discretionary account***").<br><br>• Our investment advice will cover a limited selection of investments. Other firms could provide advice on a wider range of choices, some of which might have lower costs. |

| *Broker-Dealer Services* **Brokerage Accounts** | *Investment Adviser Services* **Advisory Accounts** |
|---|---|
| **Our Obligations to You.** *We must abide by certain laws and regulations in our interactions with you.* ||
| • We must act in your best interest and not place our interests ahead of yours when we recommend an investment or an investment strategy involving securities. When we provide any service to you, we must treat you fairly and comply with a number of specific obligations. Unless we agree otherwise, we are not required to monitor your portfolio or investments on an ongoing basis.<br><br>• Our interests can conflict with your interests. When we provide recommendations, we must eliminate these conflicts or tell you about them and in some cases reduce them. | • We are held to a fiduciary standard that covers our entire investment advisory relationship with you. For example, we are required to monitor your portfolio, investment strategy and investments on an ongoing basis.<br><br>• Our interests can conflict with your interests. We must eliminate these conflicts or tell you about them in a way you can understand, so that you can decide whether or not to agree to them. |

| *Broker-Dealer Services* <br> **Brokerage Accounts** | *Investment Adviser Services* <br> **Advisory Accounts** |
|---|---|
| **Fees and Costs.** *Fees and costs affect the value of your account over time. Please ask your financial professional to give you personalized information on the fees and costs that you will pay.* ||
| • *Transaction-based fee*s. You will pay us a fee every time you buy or sell an investment. This fee, commonly referred to as a commission, is based on the specific transaction and not the value of your account. With stocks or exchange-traded funds, this fee is usually a separate commission. With other investments, such as bonds, this fee might be part of the price you pay for the investment (called a "***mark-up***" or "***mark down***"). With mutual funds, this fee (typically called a "***load***") reduces the value of your investment.<br><br>• Some investments (such as mutual funds and variable annuities) impose additional fees that will reduce the value of your investment over time. Also, with certain investments such as variable annuities, you may have to pay fees such as "***surrender charges***" to sell the investment.<br><br>• Our fees vary and are negotiable. The amount you pay will depend, for example, on how much you buy or sell, what type of investment you buy or sell, and what kind of account you have with us.<br><br>• We charge you additional fees, such as custodian fees, account maintenance fees, and account inactivity fees.<br><br>• The more transactions in your account, the more fees we charge you. We therefore have an incentive to encourage you to engage in transactions. | • *Asset-based fees.* You will pay an on-going fee at the end of each quarter based on the value of the cash and investments in your advisory account.<br><br>The amount paid to our firm and your financial professional generally does not vary based on the type of investments we select on your behalf. The asset-based fee reduces the value of your account and will be deducted from your account.<br><br>For some advisory accounts, called ***wrap fee programs***, the asset-based fee will include most transaction costs and custody services, and as a result wrap fees are typically higher than non-wrap advisory fees.<br><br>• Some investments (such as mutual funds and variable annuities) impose additional fees that will reduce the value of your investment over time. Also, with certain investments such as variable annuities, you may have to pay fees such as "***surrender charges***" to sell the investment.<br><br>• Our fees vary and are negotiable. The amount you pay will depend, for example, on the services you receive and the amount of assets in your account. |

| **Broker-Dealer Services** **Brokerage Accounts** | **Investment Adviser Services** **Advisory Accounts** |
|---|---|
| • From a cost perspective, you may prefer a transaction-based fee if you do not trade often or if you plan to buy and hold investments for longer periods of time. | • For accounts not part of the wrap fee program, you will pay a transaction fee when we buy and sell an investment for you. You will also pay fees to a broker-dealer or bank that will hold your assets (called "*custody*"). Although transaction fees are usually included in the wrap program fee, sometimes you will pay an additional transaction fee (for investments bought and sold outside the wrap fee program).<br><br>• The more assets you have in the advisory account, including cash, the more you will pay us. We therefore have an incentive to increase the assets in your account in order to increase our fees. You pay our fee quarterly even if you do not buy or sell.<br><br>• Paying for a wrap fee program could cost more than separately paying for advice and for transactions if there are infrequent trades in your account.<br><br>• An asset-based fee may cost more than a transaction-based fee, but you may prefer an asset-based fee if you want continuing advice or want someone to make investment decisions for you. You may prefer a wrap fee program if you prefer the certainty of a quarterly fee regardless of the number of transactions you have. |

| *Broker-Dealer Services* <br> **Brokerage Accounts** | *Investment Adviser Services* <br> **Advisory Accounts** |
|---|---|
| ***Additional Information.*** *We encourage you to seek out additional information.* ||

• We have legal and disciplinary events. Visit Investor.gov for a free and simple search tool to research our firm and our financial professionals.

• For additional information about our brokers and services, visit Investor.gov or BrokerCheck (BrokerCheck.Finra.org), our website ABCFinServ.com, and your account agreement. For additional information on advisory services, see our Form ADV brochure on IAPD, on Investor.gov, or on our website (ABCFinServe.com/FormADV) and any brochure supplement your financial professional provides.

• To report a problem to the SEC, visit Investor.gov or call the SEC's toll-free investor assistance line at (800)-732-0330. To report a problem to FINRA, visit www.FINRA.org/complaints. If you have a problem with your investments, account or financial professional, please contact us in writing.

| *Broker-Dealer Services* <br> **Brokerage Accounts** | *Investment Adviser Services* <br> **Advisory Accounts** |
|---|---|
| **Conflicts of Interest.** *We benefit from the services we provide to you.* ||
| • We can make extra money by selling you certain investments, such as mutual funds, either because they are managed by someone related to our firm or because they are offered by companies that pay our firm to offer their investments. Your financial professional also receives more money if you buy these investments.<br><br>• We have an incentive to offer or recommend certain investments, such as mutual funds, because the manager or sponsor of those investments shares with us revenue it earns on those investments.<br><br>• We can buy investments from you, and sell investments to you, from our own accounts (called "***acting as principal***"). We can earn a profit on these trades, so we have an incentive to encourage you to trade with us. | • We can make extra money by advising you to invest in certain investments, such as mutual funds, because they are managed by someone related to our firm. Your financial professional also receives more money if you buy these investments.<br><br>• We have an incentive to advise you to invest in certain investments, such as mutual funds, because the manager or sponsor of those investments shares with us revenue it earns on those investments.<br><br>• We can buy investments from you, and sell investments to you, from our own accounts (called "***acting as principal***"), ***but only with your specific approval on each transaction.*** We can earn a profit on these trades, so we have an incentive to encourage you to trade with us. |

| *Broker-Dealer Services* **Brokerage Accounts** | *Investment Adviser Services* **Advisory Accounts** |
|---|---|
| *Key Questions to Ask.* Ask our financial professionals these key questions about our investment services and accounts. ||

1. Given my financial situation, why should I choose an advisory account? Why should I choose a brokerage account?

2. Do the math for me. How much would I expect to pay per year for an advisory account? How much for a typical brokerage account? What would make those fees more or less? What services will I receive for those fees?

3. What additional costs should I expect in connection with my account?

4. Tell me how you and your firm make money in connection with my account. Do you or your firm receive any payments from anyone besides me in connection with my investments?

5. What are the most common conflicts of interest in your advisory and brokerage accounts? Explain how you will address those conflicts when providing services to my account.

6. How will you choose investments to recommend for my account?

7. How often will you monitor my account's performance and offer investment advice?

8. Do you or your firm have a disciplinary history? For what type of conduct?

9. What is your relevant experience, including your licenses, education, and other qualifications? Please explain what the abbreviations in your licenses are and what they mean.

10. Who is the primary contact person for my account, and is he or she a representative of an investment adviser or a broker-dealer? What can you tell me about his or her legal obligations to me? If I have concerns about how this person is treating me, who can I talk to?

Appendix B. The shorter version of Form CRS developed by Kleimann Communication Group with support from AARP (participants read each section on a separate page)

|  | **Investment Advisor Services** | **Broker-Dealer Services** |
|---|---|---|
| **What is different in our relationship with you?** | **INVESTMENT ADVICE AND MANAGEMENT**<br>**This is an advisory relationship.**<br>• We offer personalized investment advice.<br>• We have an ongoing advisory relationship of trust and confidence with you. | **SALES RECOMMENDATIONS AND TRADING**<br>**This is a sales relationship.**<br>• We offer brokerage services.<br>• We have a sales-based transactional relationship with you. |

|  | **Investment Advisor Services** | **Broker-Dealer Services** |
|---|---|---|
| **What is different in how you pay for our services?** | **ADVISORY FEES**<br>**You pay ongoing fees for advice and implementation based on a percentage of the value of the assets in your account.**<br>• Alternatively, you may contract with us to provide limited services, for which you will pay a flat or hourly fee.<br>• You may pay additional fees.<br><br>Visit www.samplefees.com to see a list of typical fees we charge. You may also request a paper copy | **COMMISSIONS AND OTHER FEES**<br>**You pay a commission or other sales fee for each transaction in your account based on the size of the transaction and the product purchased.**<br>• You may pay ongoing asset-based fees for some products in your account that we recommend.<br>• You may pay additional fees.<br><br>Visit www.samplefees.com to see a list of typical fees we charge. You may also request a paper copy |

|  | Investment Advisor Services | Broker-Dealer Services |
|---|---|---|
| **What is different in the services we provide?** | **For investments, we**<br>• work with you to identify your investment goals and develop strategies to meet those goals;<br>• analyze your entire portfolio and profile, including your age and the time you have to meet your goals; and<br>• select investments to meet your goals.<br><br>**For monitoring and oversight, we**<br>• monitor your investments over time to ensure they meet your investment goals;<br>• provide quarterly account statements; and<br>• meet with you at least annually to discuss your investment progress and any changes to your goals and profile.<br><br>**Other options**<br>You may choose how much you want to be involved in overseeing your investments. See www.involvementoptions.com for | **For investments, we**<br>• consider your investment goals and profile, including your age and the time you have to meet those goals;<br>• make buy and sell recommendations to you; and<br>• execute the transactions **you decide** to make based on our recommendations.<br><br>**For monitoring and oversight, we**<br>• do **NOT** monitor your account after the transaction.<br>• do provide quarterly account statements.<br><br>**YOU make the final decisions on the transactions, and YOU monitor your own transactions and portfolio.**<br><br>**Other options**<br>You may choose an account in which you make investment decisions and trade on your own with no Broker-Dealer recommendations. See www.accountoptions.com for more information on this choice. |

|  | **Investment Advisor Services** | **Broker-Dealer Services** |
|---|---|---|
| **What is different in our legal obligation to you?** | **In our advisory relationship, we must follow the highest *legal* standard of conduct, called a *fiduciary* standard.**<br><br>**We are required to**<br>• put your financial interests ahead of our own;<br>• review your financial situation and profile as we select investments to meet your goals;<br>• disclose and get your consent for any conflicts of interest (see next section); and<br>• monitor your account continuously throughout our relationship.<br><br>**We are required to follow these legal obligations**<br>• for ALL advice we provide to you, not just when making investment recommendations; and<br>• for the entire length and scope of our advisory relationship<br><br>**We are *not* required to:**<br>choose the lowest cost, least risky, or best performing product. | **In our sales relationship, we must follow a *best interest* standard.**<br><br>**We are required to**<br>• review your financial situation and profile as we recommend investments we have available; and<br>• disclose material information about the investments we recommend and mitigate any conflicts of interest (see next section).<br><br>**We are required to follow these legal obligations**<br>• ONLY when making sales recommendations.<br><br>**We are *not* required to:**<br>• follow this legal obligation for anything other than our sales recommendations;<br>• monitor your account unless you contract separately for that service; or<br>• choose the lowest cost, least risky, or best performing product. |

|  | Investment Advisor Services | Broker-Dealer Services |
|---|---|---|
| **What is different in how we handle a financial conflict of interest?** | As part of our legal obligations, we must identify and disclose when our interests conflict with yours, and *you must consent to the conflict either directly or indirectly.*<br><br>**Sometimes our interests conflict with yours.** This means investment advice that results in extra income for us is not the best for you. For example:<br>• **Because we receive asset-based fees,** we may try to maximize the amount of money you invest with us.<br>• **Because we receive payments from other companies,** we may recommend their investments, even if other options have lower costs or better performance.<br><br>See www.conflictspolicy.com where we explain how our firm meets our legal obligation to protect your best interests. | As part of our legal obligations, we must identify, disclose, and in some cases *mitigate a situation when our interests conflict with yours.*<br><br>**Sometimes our interests conflict with yours.** This means a sales recommendation that results in extra income for us is not the best for you. For example:<br>• **Because we get paid only when you complete a transaction,** we may encourage you to trade more often.<br>• **Because we receive payments from other companies,** we may recommend their investments, even if other options have lower costs or better performance.<br>• **Because we get higher commissions from some products,** we may encourage you to buy those products, even if other options are better for you.<br><br>See www.conflictspolicy.com where we explain how our firm meets our legal obligation to protect your best interests. |
| **To research our firm** | Visit investor.gov for a free, simple resource on investing and for links to search tools to research whether our firm or any of our financial professionals has a disciplinary record. ||
| **Alert!** | Many financial professionals use titles such as Financial Consultant, Wealth Manager, Adviser, or variations of that title. A firm or individual's title does NOT mean that they are acting in an advisory capacity. Every time you interact, ask your financial professional this important question: Are you working in an advisory or selling role? ||

Appendix C. Preferences for investment features

|   | Feature A | Feature B |
|---|-----------|-----------|
| 1 | I prefer making my own investment decisions | I prefer having my financial professionals make my investment decisions for me |
| 2 | I prefer paying a commission per trade | I prefer paying a fee based on a percentage of my assets |
| 3 | I prefer monitoring my own accounts | I prefer having my financial professionals monitor my accounts for me |

Appendix D. Answers to the comprehension questions were derived based on direct quotes from Form CRS

|  | Longer Version | | Shorter Version | |
|---|---|---|---|---|
| Questions | Investment Adviser | Broker-dealer | Investment Adviser | Broker-dealer |
| 1 | "If you open an advisory account, you will pay an on-going *asset-based fee* for our services." | "If you open a brokerage account, you will pay us a *transaction-based fee*, generally referred to as a commission, every time you buy or sell an investment." | "We have an ongoing advisory relationship of trust and confidence with you." | "We have a sales-based transactional relationship with you." |
| 2 | "We may give you advice and you decide what investments to buy and sell." | "We may recommend investments for your account, but the ultimate investment decision...will be yours." | "We select investments to meet your goals...You may choose how much you want to be involved in overseeing your investments." | "YOU make the final decisions on the transactions" |
| 3 | "We are held to a fiduciary standard that covers our entire investment advisory relationship with you." | "We must act in your best interest and not place our interests ahead of yours when we recommend an investment or an investment strategy involving securities." | "In our advisory relationship, we must follow the highest *legal* standard of conduct, called a *fiduciary* standard." | "In our sales relationship, we must follow a *best interest* standard." |
| 4 | "*Asset-based fees*. You will pay an on-going fee at the end of each quarter based on the value of the cash and investments in your advisory account" | "*Transaction-based fees*. You will pay us a fee every time you buy or sell an investment. This fee, commonly referred to as a commission, is based on the specific transaction and not the value of your account." | "You pay ongoing fees for advice and implementation based on a percentage of the value of the assets in your account." | "You pay a commission or other sales fee for each transaction in your account based on the size of the transaction and the product purchased." |
| 5 | "*Asset-based fees*. You will pay an on-going fee at the end of each quarter based on the value of the cash and investments in your advisory account" | "*Transaction-based fees*. You will pay us a fee every time you buy or sell an investment. This fee, commonly referred to as a commission, is based on the specific transaction and not the value of your account." | "You pay ongoing fees for advice and implementation based on a percentage of the value of the assets in your account." | "You pay a commission or other sales fee for each transaction in your account based on the size of the transaction and the product purchased." |
| 6 | "We can make extra money by advising you to invest in certain investments, such as mutual funds, because they are managed by someone related to our firm." | "We can make extra money by selling you certain investments, such as mutual funds, either because they are managed by someone related to our firm or because they are offered by companies that pay our firm to offer their investments." | "Because we receive payments from other companies, we may recommend their investments, even if other options have lower costs or better performance." | "Because we receive payments from other companies, we may recommend their investments, even if other options have lower costs or better performance. Because we get higher commissions from some products, we may encourage you to buy those products, even if other options are better for you." |
| 7 | "We can buy | "We can buy | "Because we receive | "Because we get paid |
|   | investments from you, and sell investments to you, from our own accounts (called "*acting as principal*"), *but only with your specific approval on each transaction*. We can earn a profit on these trades, so we have an incentive to encourage you to trade with us." | investments from you, and sell investments to you, from our own accounts (called "*acting as principal*"). We can earn a profit on these trades, so we have an incentive to encourage you to trade with us." | asset-based fees, we may try to maximize the amount of money you invest with us." | only when you complete a transaction, we may encourage you to trade more often." |
| 8* | "We will...regularly monitor your account." | "We can offer you additional services to assist you in...monitoring the performance of your account but you might pay more." | "We monitor your investments over time to ensure they meet your investment goals." | "We do NOT monitor your account after the transaction. and YOU monitor your own transactions and portfolio." |

*For question 8, the longer version and the shorter version disagree regarding whether broker-dealers can monitor a client's account. "Both" was coded as the answer for the longer version, and "Investment advisor" the answer for the shorter version.

Appendix E. Belief distribution for question 8: account monitoring

| Version | Baseline M(SD) | Post Disclosure M(SD) |
|---|---|---|
| Shorter | 65.17(26.39) | 76.85(23.21)*** |
| Longer | 66.48(26.53) | 85.24(22.07)*** |

Note: Belief was measured on a 0-100 scale; 0 = definitely broker-dealers, 50 = both, and 100 = definitely investment advisors. Paired sample t tests results are shown; $p < .001$***.

Appendix F. An example of comprehension questions

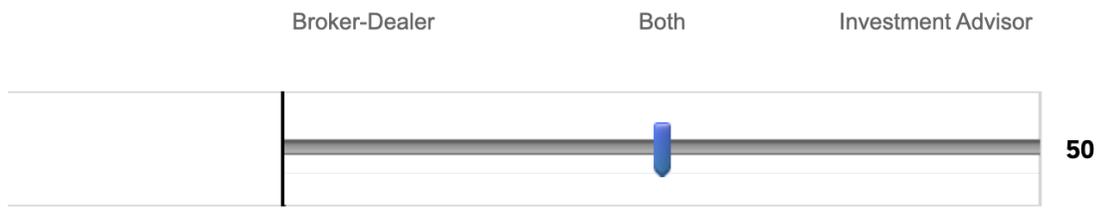

Appendix G. Decision making measurements

You have money that you want to invest and you do not currently have any investment accounts. You have contacted an investment firm that offers two types
of services: broker-dealer services and investment advisor services. Which services do you choose?

○ Broker-Dealer Services

○ Investment Advisor Services

How strong is your preference for the type of account you selected above? (drag or leave the slider in the center to indicate your response)

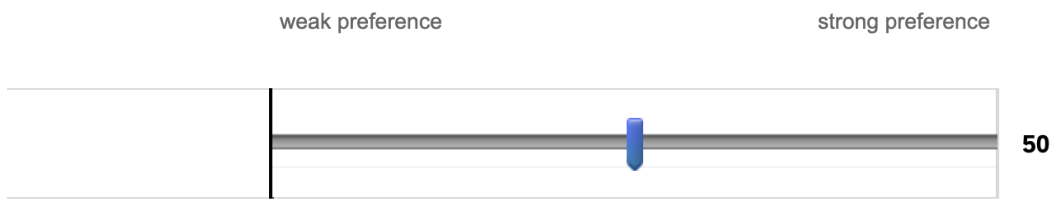

weak preference · · · strong preference · · · 50